\documentclass[3p,times,twocolumn]{elsarticle}

\usepackage{amssymb}
\usepackage{lineno}

\usepackage{xspace}

\newcommand{\snn}{\ensuremath{\sqrt{s_{NN}}}\xspace}
\newcommand{\Raa}{\ensuremath{R_\mathrm{AA}}\xspace}

\newcommand{\Nbin}{\ensuremath{N_\mathrm{bin}}\xspace}
\newcommand{\Npart}{\ensuremath{N_\mathrm{part}}\xspace}

\newcommand{\pim}{\ensuremath{\pi^-}\xspace}
\newcommand{\pip}{\ensuremath{\pi^+}\xspace}
\newcommand{\Km}{\ensuremath{\mathrm{K}^{-}}\xspace}

\newcommand{\ele}{\ensuremath{\mathrm{e}^{-}}\xspace}
\newcommand{\pos}{\ensuremath{\mathrm{e}^{+}}\xspace}
\newcommand{\nele}{\ensuremath{\nu}_e\xspace}
\newcommand{\Dz}{\ensuremath{\mathrm{D}^{0}}\xspace}
\newcommand{\Ds}{\ensuremath{\mathrm{D}^{*}}\xspace}

\newcommand{\Jpsi}{\ensuremath{\mathrm{J}/\psi}\xspace}
\newcommand{\Ups}{\ensuremath{\Upsilon}\xspace}
\newcommand{\GeV}{\ensuremath{\mathrm{GeV}}\xspace}
\newcommand{\MeV}{\ensuremath{\mathrm{MeV}}\xspace}
\newcommand{\pT}{\ensuremath{p_\mathrm{T}}\xspace}
\newcommand{\kT}{\ensuremath{k_\mathrm{T}}\xspace}

\newcommand{\cc}{\ensuremath{c\bar{c}}\xspace}

\usepackage{color}

\biboptions{sort&compress}

\begin{document}

\begin{frontmatter}




\title{Heavy Flavor Measurements at STAR}


\author{R\'obert V\'ertesi for the STAR Collaboration}
\address{Nuclear Physics Institute ASCR, 25068
\v{R}e\v{z}, Czech Republic}

\address{}

\begin{abstract}
We present a selection of recent heavy flavor results from the STAR experiment. 
Measurements of \Dz and \Ds meson production in $\sqrt{s}$=200 and
500 GeV p+p, as well as in \snn{}=200 GeV d+Au, Au+Au and 193 \GeV U+U collisions are presented and
implications on the production mechanism are discussed. 
We report on the production and elliptic flow of electrons from
semi-leptonic decays of heavy flavor hadrons in \snn{}=39, 62.4
and 200 GeV Au+Au collisions. Nuclear modification of $J/\psi$
production in \snn{}=39, 62.4 and 200 GeV Au+Au, and 193 GeV U+U, and of \Ups
in 200 GeV d+Au, Au+Au, and 193 GeV U+U collisions are compared to theoretical models. Finally we
discuss the prospects of heavy flavor measurements with the recent detector upgrades.
\end{abstract}


\end{frontmatter}


\section{Introduction}
\label{sec:intro}

In ultrarelativistic heavy ion collisions, a phase transition occurs
from hadronic matter into a state of deconfined quarks and
gluons~\cite{Adams:2005dq,Adare:2008ab}. Properties of this latter state of matter, dubbed as the
strongly interacting Quark Gluon Plasma (sQGP), have been a subject of
extensive measurements at the Relativistic Heavy Ion Collider (RHIC)
in the past decade and a half. The RHIC Beam Energy Scan (BES) program Phase I.\
was dedicated to the search of turn-off signatures of the sQGP~\cite{Aggarwal:2010cw}. On the other hand, significantly improved luminosity of RHIC
in the recent years, paired with continuous detector development allows one to turn
to rare probes, such as heavy flavor production, which are
complementary to observables of light hadrons and provide us with a
deeper understanding of the strong interaction.

\subsection{Heavy flavor at RHIC energies}
Charm and bottom quarks are produced in hard QCD processes early in
the interaction, and, due to their large masses, their number is
virtually unaffected in the later stages of the reaction. Heavy
flavor quarks therefore provide a unique means of exploring the
properties of the sQGP. Open heavy flavor yields at different momenta are sensitive to
the energy loss mechanism of partons. Azimuthal anisotropy
measurements may supply us with additional information about the degree of thermalization of the medium. 

Quarkonium states are expected to be subject to sequential melting due
to the screening of the $q\bar{q}$ potential in the sQGP, and provide access to
thermodynamical properties of the medium~\cite{Mocsy:2007jz}.
Production of charmonia is abundant and therefore it is relatively easy
to measure them with precision. However, several effects
concurrent to sequential melting, such as feed-down,
recombination in the sQGP and co-mover absorption in the hadronic
phase, influence measured yields~\cite{Gorenstein:2000ck}. The interplay between these effects can be understood by
comprehensive measurements at different energies and in different colliding
systems. Bottomonium measurements are of special interest because,
contrary to charm quarks, the effect of bottom pair recombination and co-mover
absorption is negligible at RHIC energies~\cite{Rapp:2008tf}.

\subsection{Experiment}
\label{sec:exp}

The STAR detector at RHIC is a compound experiment of several
subsystems that provides a full azimuthal coverage at
mid-rapidity ($|\eta|${}$<$1). In the analyses discussed below,
momentum measurement of charged particles, as well as particle
identification based on energy loss via ionization, is done
using the Time Projection Chamber (TPC). Charged particle identification is aided by
the Time of Flight detector (TOF) at low \pT. The Barrel Electromagnetic Calorimeter
(BEMC) is used for energy measurement and further identification of
electrons. The Vertex Position Detector (VPD) provides trigger for the minimum
bias (MB) data. A detailed description of STAR is in
Ref.~\cite{Ackermann:2002ad}. 
Results of Refs.~\cite{Adamczyk:2012af,Adamczyk:2014uip,Adamczyk:2013tvk,Adamczyk:2012ey,Adamczyk:2012pw,Adamczyk:2014yew,Adamczyk:2013poh} are summarized here along with some
recent preliminary results.

\section{Open heavy flavor measurements}
\label{sec:openhf}

Currently, STAR measures open heavy flavor through two different channels. One is the direct
reconstruction of charmed mesons from their hadronic decays ($\Dz${}$\rightarrow${}$\Km\pip$ or ${\Ds}^{+}${}$\rightarrow${}$\Dz\pip${}$\rightarrow${}$\Km\pim\pip$ and their charge conjugated
counterparts)~\cite{Adamczyk:2012af,Adamczyk:2014uip}.
 \Dz and \Ds meson production was measured in $\sqrt{s}$=200 and
500 GeV p+p, as well as \Dz in $\snn$=200 GeV Au+Au and $\snn$=193 GeV
U+U MB collisions.
The invariant yields are determined in the following way: the
invariant mass peak is reconstructed from the decay products, the
  combinatorial background is subtracted using event mixing (\Dz) or swapped-sign
background (\Ds), and the residual background is removed with a
sideband fit. The raw spectrum is then corrected for acceptance and efficiency. 
This method provides direct access to the kinematics of the charmed meson. However, such
events are difficult to trigger on, and without secondary vertex
detection capabilities, the combinatorial background level is high.

 The other way is to measure {\it non-photonic electrons} (NPE) from
semileptonic decays of charm and bottom~\cite{Adamczyk:2014yew}, eg.\
$\Dz${}$\rightarrow${}$\pos\nele\Km$. Such processes
typically have branching ratios two times larger than hadronic
  channels, and high-\pT electrons are easily triggered on in the
BEMC using a {\it high tower} (HT) trigger.
It is, however, not possible to directly separate charm and
bottom contribution without secondary vertex reconstruction.
The main background for this process is the so-called photonic
electrons (PE), \ele{}\pos pairs from decays of light hadrons and from photon conversion in
detector material. The PE contribution is determined from data.

\subsection{\Dz and \Ds production}
\label{sec:d}

\begin{figure}[t]
\centering
\includegraphics[width=0.95\columnwidth]{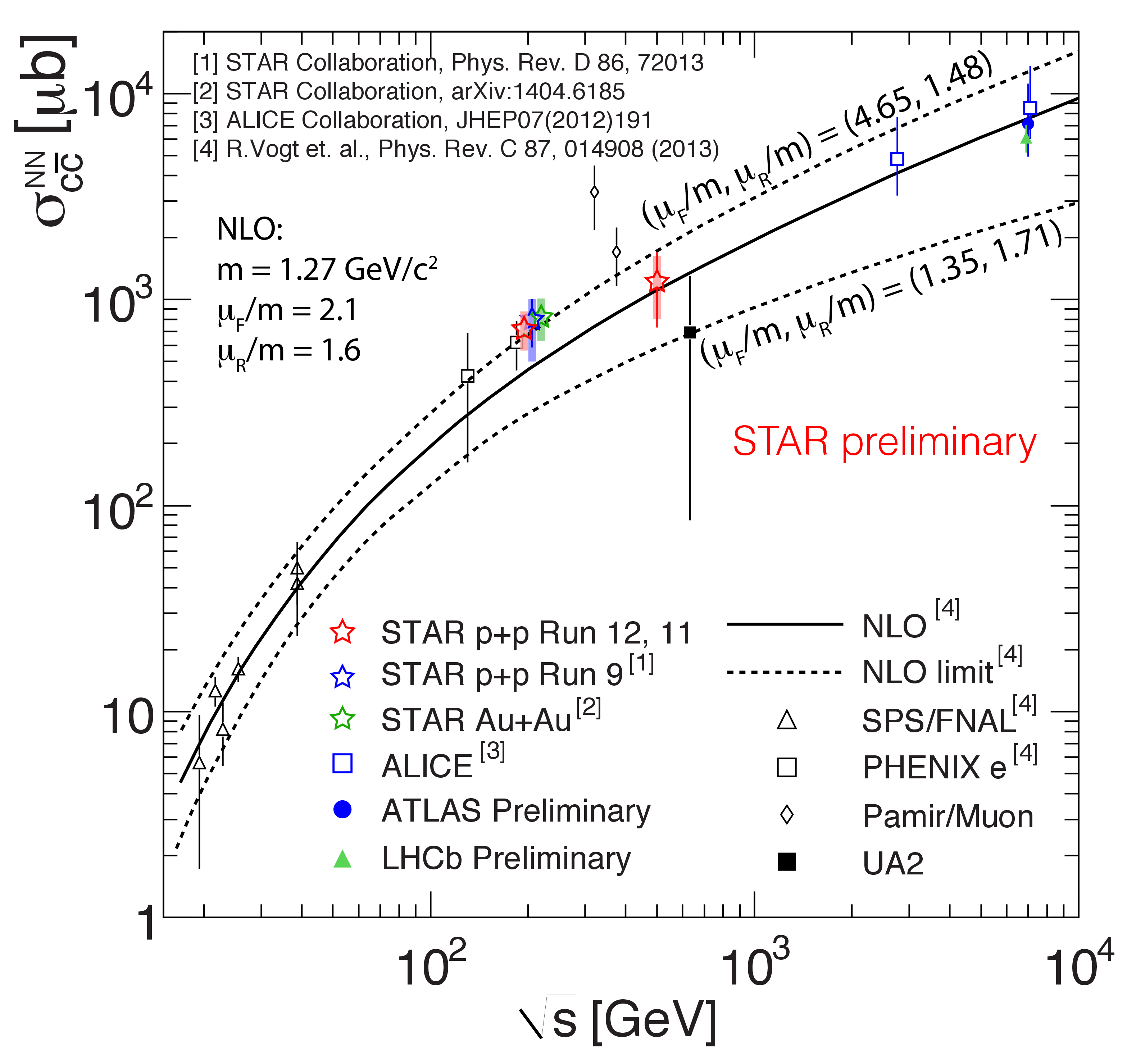}
\caption{Charm production cross sections in p+p collisions
  versus collision energy $\sqrt{s}$, compared to world data and pQCD calculations~\cite{Nelson:2012bc}.}
\label{fig:D_pp}
\end{figure}

Precision p+p measurements are essential as a benchmark for
  theoretical calculations as well as a baseline for nuclear modification.
Uncertainty in extrapolation of the \Dz and \Ds spectra towards lower
momenta used to be one of the main sources of the systematic
uncertainty on the total production cross
section~\cite{Adamczyk:2012af}. Recent high-statistics
measurements at $\sqrt{s}$=200 GeV constrained the shape of the spectrum by extending the range with a
new 0$<$\pT{}$<$0.7 GeV/$c$ point.
Fig.~\ref{fig:D_pp} shows previously published charm production cross sections
$d\sigma^{NN}_{c\bar{c}}$ at \snn{}=200 GeV,
 as well as new 200 and 500 GeV data. STAR data
points follow the trend of world data and are described by pQCD FONLL
calculations~\cite{Nelson:2012bc}. 

\begin{figure}[t]
\centering
\includegraphics[width=0.95\columnwidth]{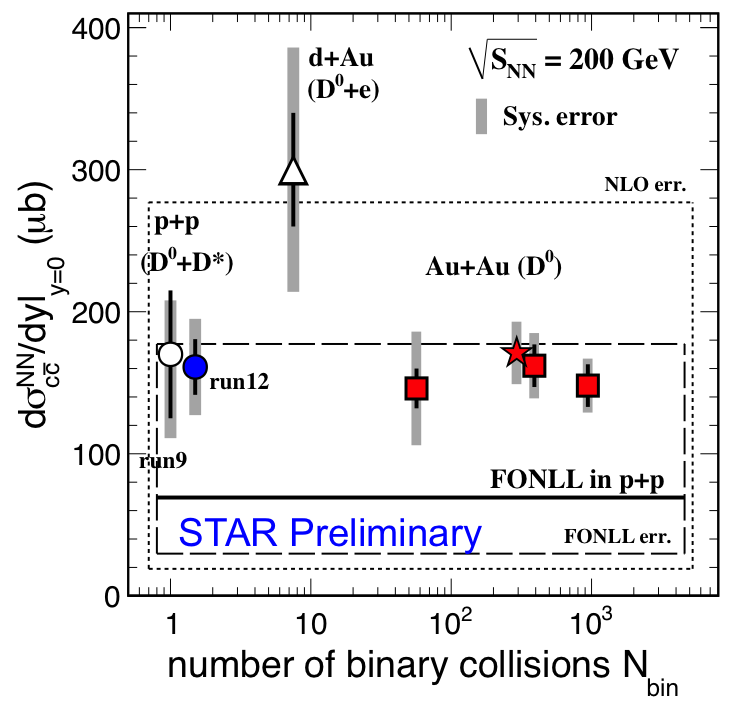}\hspace{0.05\columnwidth}
\caption{The STAR results at $\sqrt{s}$=200 and 500
  GeV are compared to different experiments and an NLO model. {\it
    Right:} Charm production cross section in STAR p+p, d+Au and Au+Au
collisions versus the number of binary collisions \Nbin.}
\label{fig:D_ppAA}
\end{figure}

\begin{figure}[t]
\centering
\includegraphics[width=\columnwidth]{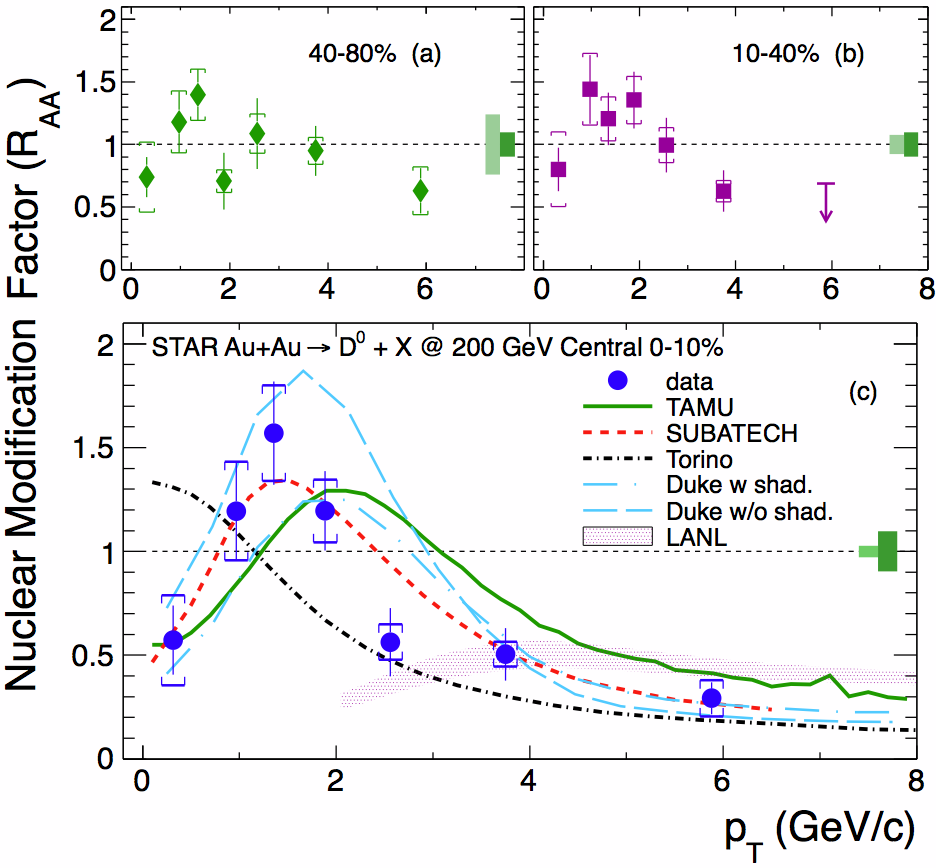}%
\caption{Nuclear modification factor of \Dz in Au+Au collisions versus
\pT in (a) peripheral, (b) mid-central and  (c) central collisions~\cite{Adamczyk:2014uip},
the latter compared to several popular models~\cite{Sharma:2009hn,He:2011qa,Gossiaux:2010yx,Cao:2013ita,Alberico:2011zy}.}
\label{fig:D0_Raa}
\end{figure}

Fig.~\ref{fig:D_ppAA} shows a comparison of
$d\sigma^{NN}_{c\bar{c}}$ in $\snn$=200 GeV p+p, d+Au and Au+Au collisions. The total
cross section follows scaling with the number of binary collisions (\Nbin), supporting the picture that charm
is produced early, in perturbative processes.
Accordingly, nuclear modification factor \Raa of the \Dz
mesons, in Fig.~\ref{fig:D0_Raa} (a), exhibit no significant modification 
in peripheral collisions at \snn{}=200 GeV~\cite{Adamczyk:2014uip}. In more central collisions, however, a
different structure emerges: while higher-\pT{} \Dz \Raa shows similar suppression
to that of light mesons~\cite{Abelev:2007ra, Agakishiev:2011dc}, a
characteristic hump-shaped enhancement
appears in the 1$<$\pT{}$<$2 GeV/$c$ range. The 10\% most central data
are compared to several models in Fig.~\ref{fig:D0_Raa} (c). High-\pT
suppression of the charmed mesons indicates a strong charm--medium
interaction~\cite{Sharma:2009hn}. The low-momentum enhancement can be
understood by models that include charm--light quark
coalescence~\cite{He:2011qa,Gossiaux:2010yx,Cao:2013ita} and allow for
charm to pick up radial flow. A model prediction without charm--light quark
coalescence~\cite{Alberico:2011zy} differs significantly from
observations in the low-momentum range~\cite{Adamczyk:2014uip}.
Model calculations of Ref.~\cite{Cao:2013ita} hint that {\it cold nuclear matter}
(CNM) effects play an important role.

\begin{figure}[t]
\includegraphics[width=\columnwidth]{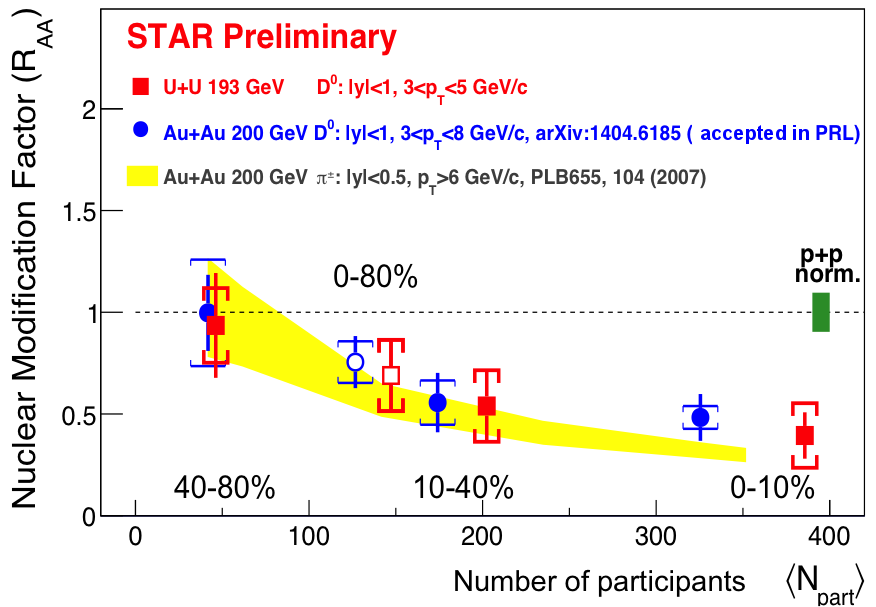}%
\caption{Nuclear modification factor of \pT{}$>$3 GeV/$c$ \Dz mesons in Au+Au (dots)
  and in preliminary U+U measurements (squares) for minimum bias collisions (open
  symbols) as well as at different \Npart values (solid symbols).}
\label{fig:D0_RaaNpart}
\end{figure}

Uranium ions are heavier than gold and have a prolate
shape, and therefore they extend the STAR measurements up to higher
number of participant nucleons \Npart. It is estimated~\cite{Kikola:2011zz} 
 that, on average, an approximately 20\% higher Bjorken energy density is achieved
 in central U+U than in central Au+Au collisions.
The \Raa of \Dz mesons with \pT{}$>$3 GeV/$c$  
in top-energy Au+Au and U+U versus \Npart is shown in Fig.~\ref{fig:D0_RaaNpart}. The
trend of increasing suppression with \Npart, observed in Au+Au
collisions, is continued in the U+U data. This trend is completely consistent
with high-\pT light mesons~\cite{Abelev:2007ra}.


\subsection{Non-photonic electrons at different energies}
\label{sec:npe}

\begin{figure}[t]
\includegraphics[width=\columnwidth]{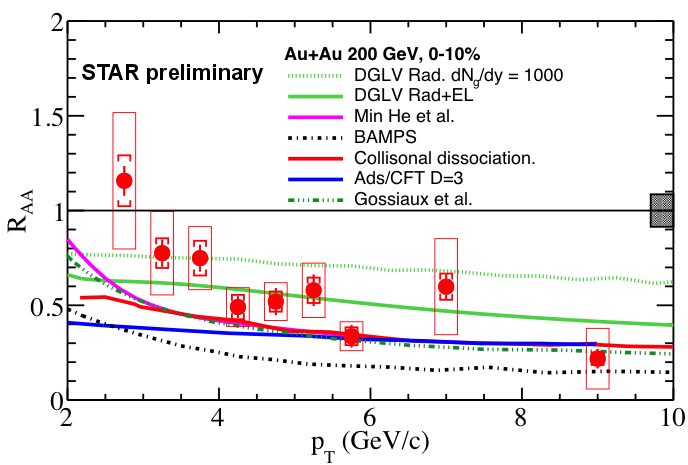}%
\caption{Non-photonic electron \Raa in central \snn{}=200 GeV Au+Au
  collisions compared to several models~\cite{Djordjevic:2005db,Buzzatti:2012pe,vanHees:2007me,Sharma:2009hn,Horowitz:2010dm}.}
\label{fig:NPE200_Raa}
\end{figure}

Fig.~\ref{fig:NPE200_Raa} shows a significant suppression 
of non-photonic electron yield in central \snn{}=200 GeV Au+Au collisions,
that is similar to light mesons. 
The results are compared to several energy loss mechanism
models~\cite{Djordjevic:2005db,Buzzatti:2012pe,vanHees:2007me,Sharma:2009hn,Horowitz:2010dm}. 
Although gluon radiation scenario~\cite{Djordjevic:2005db} is quite
successful in describing energy loss
of light hadrons, it appears not to be enough alone to explain the observed high-\pT NPE suppression.
The azimuthal anisotropy parameter $v_2$ (also called elliptic flow) is found to be
non-zero, and significantly above the non-flow estimation at low
momenta~\cite{Adamczyk:2014yew}. It is to be noted that those models
that are in better agreement with the
NPE \Raa come short in describing $v_2$ at the same time.

\begin{figure}[t]
\includegraphics[width=\columnwidth]{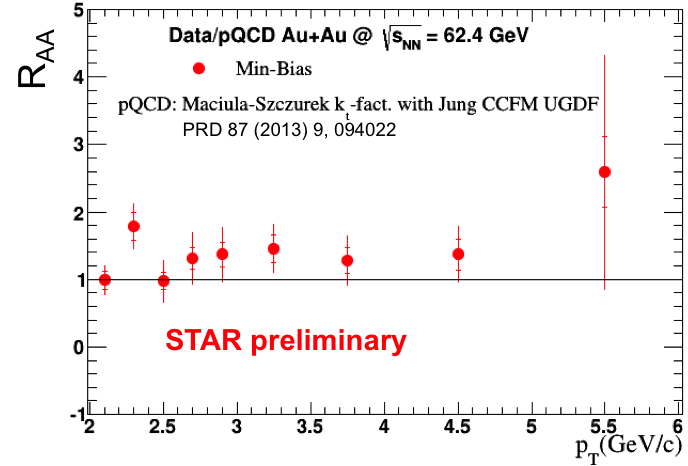}%
\caption{Non-photonic electron \Raa in \snn{}=62.4 GeV Au+Au collisions.
The p+p reference used here is from pQCD calculations with \kT-factorization
technique~\cite{Maciula:2013kd}.}
\label{fig:NPE62_Raa}
\end{figure}

\begin{figure}[t]
\includegraphics[width=\columnwidth]{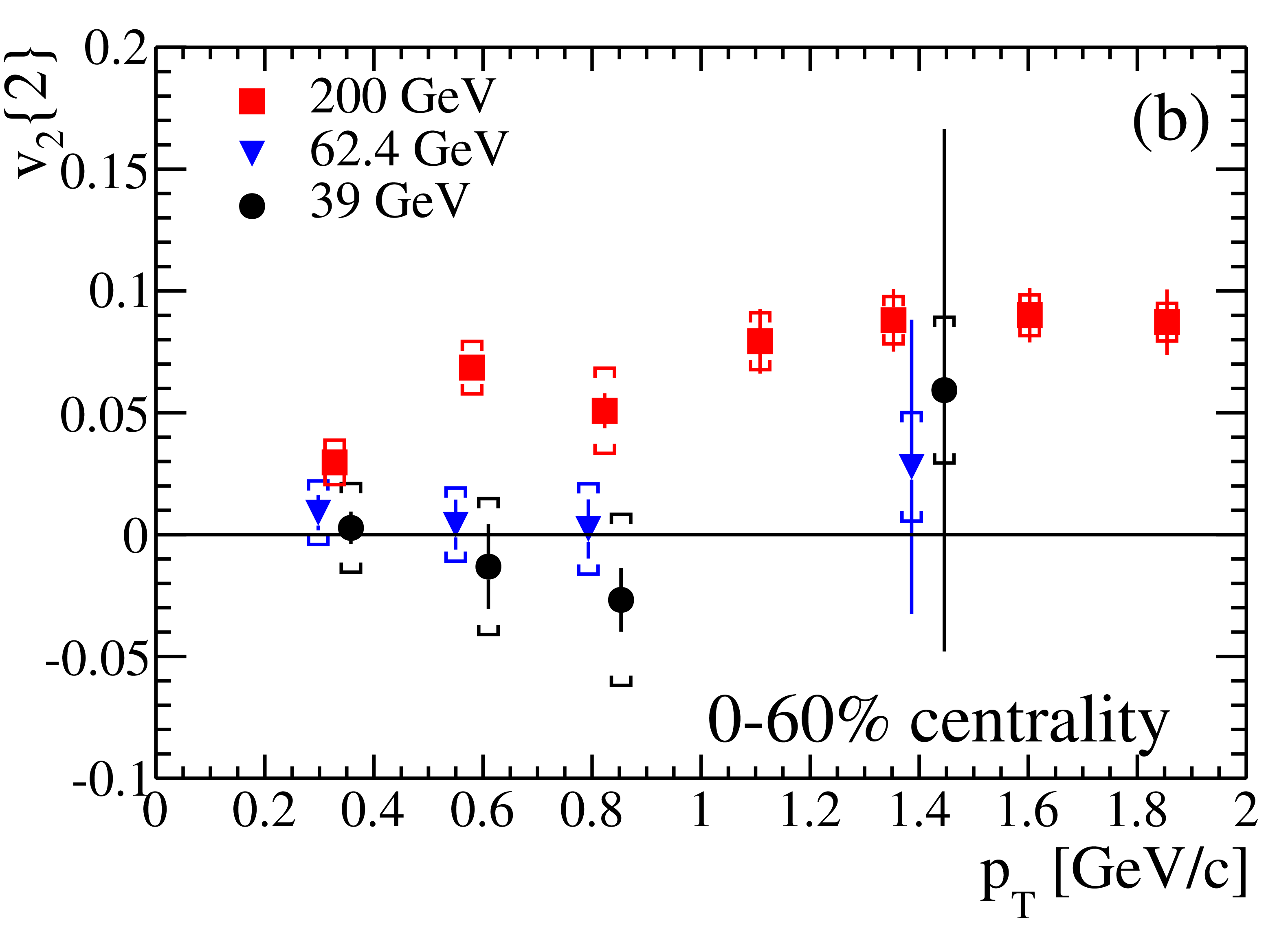}%
\caption{Non-photonic electron $v_2$ in \snn=39, 62.4 and 200 GeV Au+Au collisions~\cite{Adamczyk:2014yew}.}
\label{fig:NPE_v2}
\end{figure}

The picture emerging from recent lower energy measurements is,
however, radically different from top RHIC energy. Fig.~\ref{fig:NPE62_Raa}
shows the \Raa of NPE in \snn{}=62.4 GeV Au+Au collisions. The p+p
reference used here is from pQCD calculations with \kT-factorization
technique~\cite{Maciula:2013kd}, which gives an upper limit
for NPE production in p+p collisions~\cite{Basile:1981dn}. 
The \Raa is generally flat, showing no NPE suppression in \snn{}=62.4 GeV Au+Au collisions. 
Fig.~\ref{fig:NPE_v2} shows the NPE elliptic flow in
\snn=39, 62.4 and 200 GeV Au+Au collisions. While at top RHIC energy
one observes substantial flow, $v_2$ is consistent with zero in the
case of the two lower energies, and differs significantly from 
top energy results in the 0.5$<$\pT{}$<$1 GeV/$c$ range.
These results suggest that the effect of the thermalized, hot matter does not
dominate heavy flavor production anymore at 62.4 GeV and below.

\section{Quarkonium results}
\label{sec:onium}

Quarkonia are reconstructed via the dielectron decay channel
($\Jpsi${}$\rightarrow${}$\ele\pos$, $\Ups${}$\rightarrow${}$\ele\pos$).
Minimum bias data were used in the BES \Jpsi analyses~\cite{Adamczyk:2013tvk}, with an event
mixing background subtraction. At \snn{}=200 GeV, both MB and HT
triggered data were used to maximize luminosity at high-\pT~\cite{Adamczyk:2012ey}. 
HT triggered data were used in the \Ups
analyses~\cite{Adamczyk:2013poh}. Like-sign pair combinatorial background
subtraction was applied.
There is a substantial background contribution
from Drell-Yan and open ${b\bar{b}}$ processes, which are accounted
for using a simultaneous fit with templates of pre-set shapes from
models, together with the signal. 

\subsection{Suppression and flow of the \Jpsi}
\label{sec:jpsi}

\begin{figure}[t]
\vspace{-2mm}
\includegraphics[width=\columnwidth]{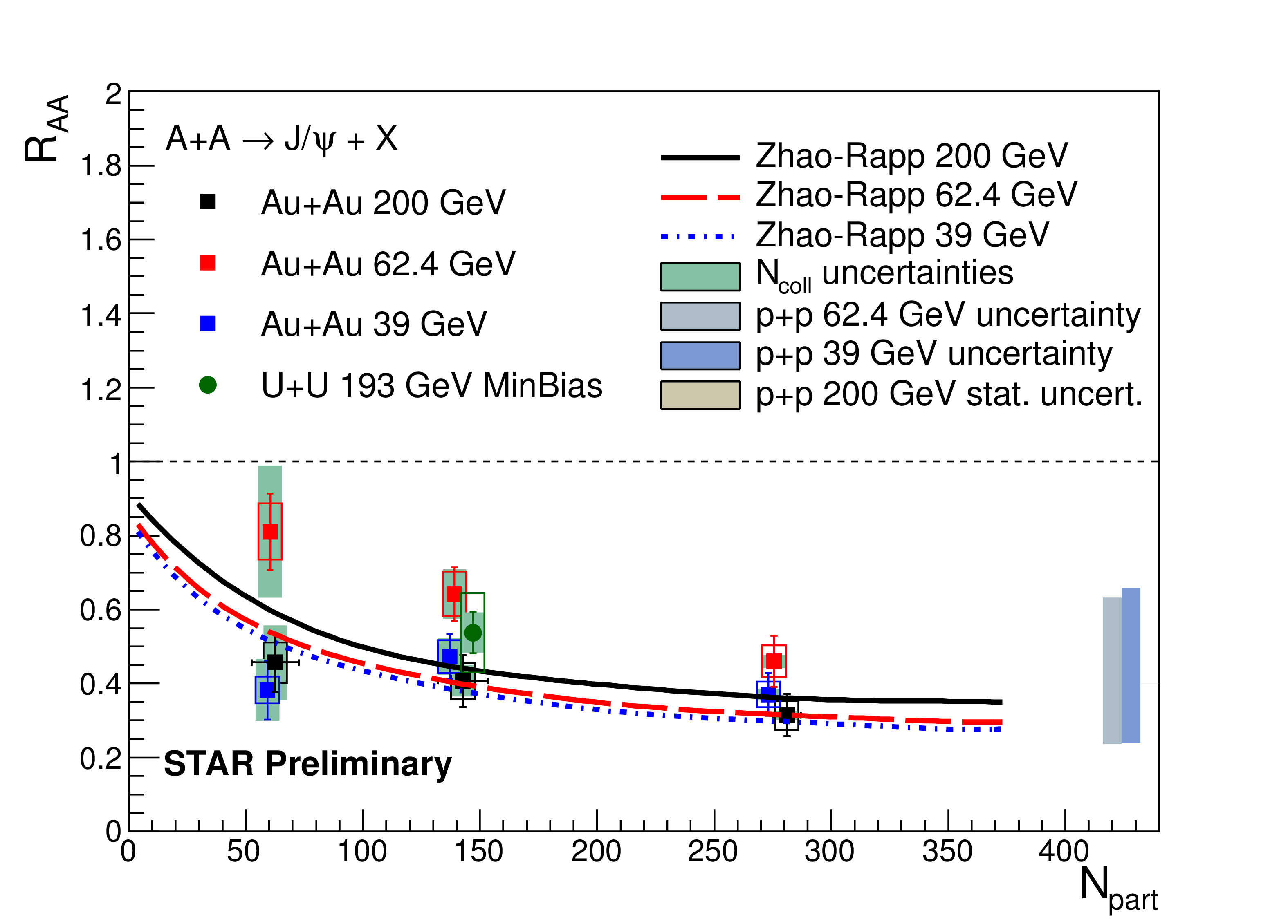}%
\caption{Low-momentum (\pT{}$<$5 GeV/$c$) \Jpsi nuclear modification factor
  vs.\ the number of participants in \snn=39, 62.4 and 200 GeV Au+Au and
  194 GeV U+U collisions (black, red, blue squares and a green dot, respectively),
compared to a model calculation~\cite{Zhao:2010nk}.}
\label{fig:JpsiBES_Raa}
\end{figure}

The \Jpsi nuclear modification factor is shown in
Fig.~\ref{fig:JpsiBES_Raa} for \snn=39, 62.4 and 200 GeV Au+Au
collisions as a function of \Npart. Minimum bias
U+U data at \snn=193 GeV is also shown on the plot. Due to the lack of p+p data with
sufficient statistics, Color Evaporation Model (CEM) predictions~\cite{Nelson:2012bc}
are used as 39 and 62.4 GeV references. A significant suppression is observed at all energies in mid-peripheral to central collisions,
and all four datasets are consistent with each other within
uncertainties. Model predictions from Zhao and Rapp~\cite{Zhao:2010nk} are consistent with
data. This model includes in-medium dissociation of the \Jpsi as well as later
regeneration from \cc pairs. 
The weak dependence on collision energy predicted by the model and seen in the data
suggests that different contributions that modify 
the \Jpsi yield largely cancel each other in the observed \Raa.

\begin{figure}[t]
\centering
\includegraphics[width=0.95\columnwidth]{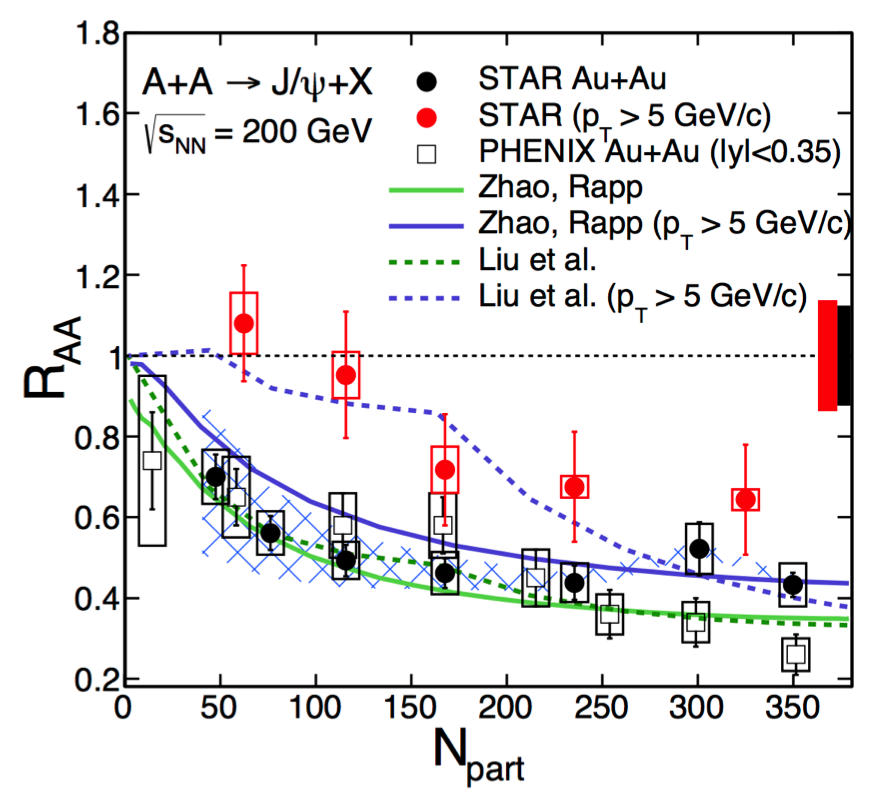}%
\caption{\Jpsi nuclear modification factor in \snn=200 GeV Au+Au
  collisions from STAR (black dots)~\cite{Adamczyk:2013tvk} and PHENIX (open circles)~\cite{Adare:2006ns}, and STAR high-momentum-only (\pT$>$5 GeV/$c$) \Jpsi \Raa (red dots)~\cite{Adamczyk:2012ey} vs.\ the number of participant nucleons, compared to model calculations~\cite{Zhao:2010nk,Liu:2009nb}.}
\label{fig:HiPtJpsi_Raa}
\end{figure}

The different contributions to the \Jpsi suppression are, however, 
momentum-dependent. High-\pT \Jpsi production is much less
affected by CNM effects as well as regeneration in the later
stages, and hot nuclear modification becomes the
dominant factor~\cite{Zhao:2010nk}. Fig.~\ref{fig:HiPtJpsi_Raa} shows the 
\Raa for high-momentum (\pT{}$>$5 GeV) \Jpsi mesons. 
In central collisions one still observes a significant suppression
that is consistent with the minimum bias data, 
while in the case of peripheral collisions it is gone and \Raa is
consistent with unity.  It is to be noted that the current STAR \Jpsi
measurements are inclusive, where the contribution from B meson
feed-down can be as high as 15--25\% above \pT{}$=$5 GeV.  The model of Ref.~\cite{Zhao:2010nk}
underpredicts the change in peripheral \Raa when going from \pT{}-integrated to high-momentum data. However, the model of Liu {\it et
  al.}~\cite{Liu:2009nb}, which includes the same main components but also considers melting of excited charmonium states feeding down to
\Jpsi, provides good description to the data. 

\begin{figure}[t]
\centering
\vspace{2mm}
\includegraphics[width=1.05\columnwidth]{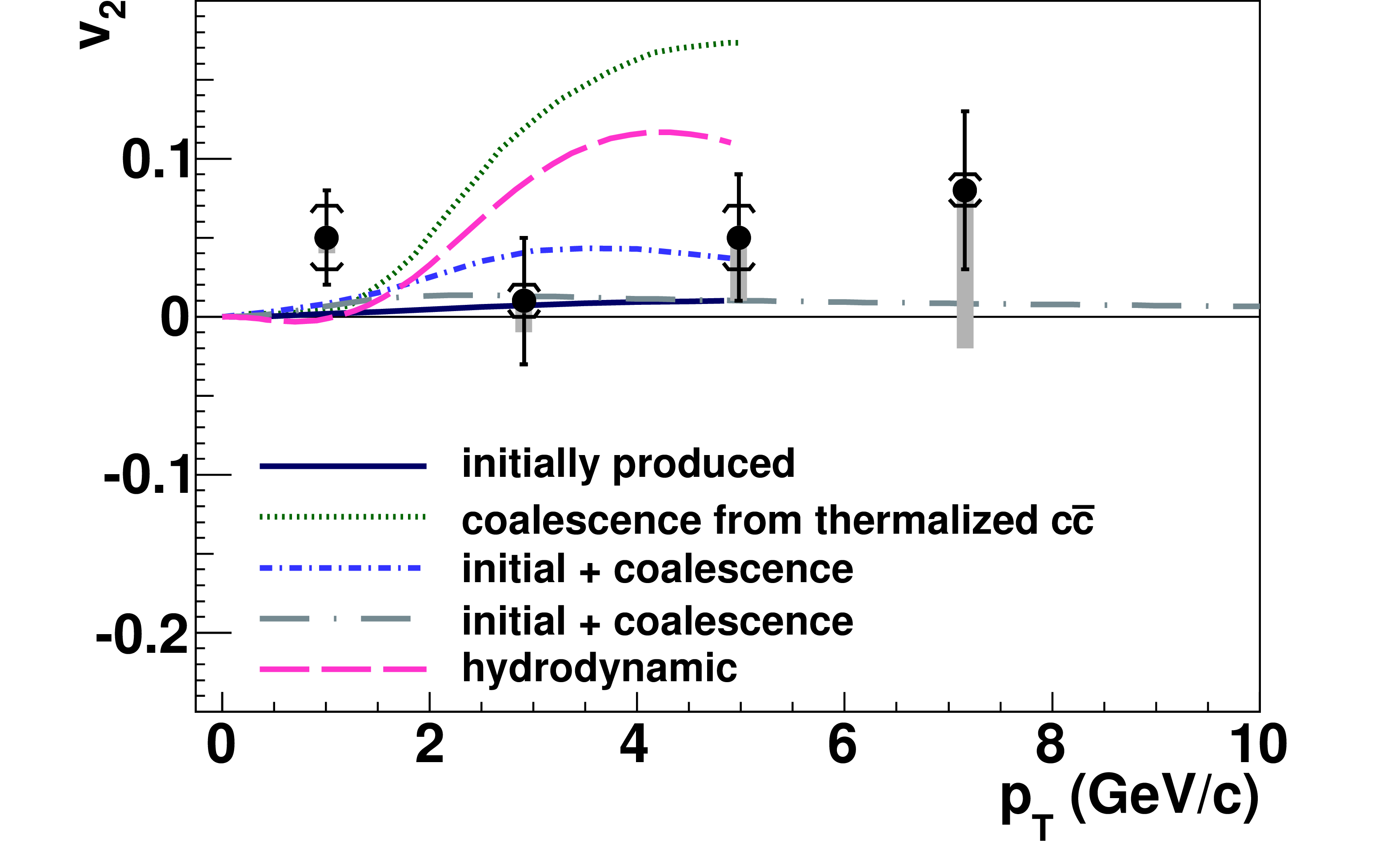}
\caption{Azimuthal anisotropy $v_2$ of the \Jpsi mesons in minimum
  bias data~\cite{Adamczyk:2012pw}, compared to model calculations (lines). The shaded boxes
  represent the possible extent of the non-flow contribution.}
\label{fig:Jpsi_v2}
\end{figure}

Those \Jpsi mesons that are created late in the reaction may be 
thermalized and thus they may take part in the collective motion that the bulk of
hadrons exhibit. Fig.~\ref{fig:Jpsi_v2}
shows the azimuthal anisotropy parameter $v_2$ of the \Jpsi mesons in minimum
bias data. Above \pT{}$>$2 GeV, STAR data shows complete consistency with the
non-flow estimation, unlike $v_2$ of light mesons and the $\phi$ mesons~\cite{Adamczyk:2012pw}.
This renders the scenario unlikely where \Jpsi mesons mainly stem
from late, thermalized coalescence of \cc pairs~\cite{Ravagli:2007xx},
and supports early production of a significant fraction of the \Jpsi mesons~\cite{Zhao:2008vu,Liu:2009gx}.

\subsection{Suppression of the \Ups states}
\label{sec:ups}

\begin{figure}[t]
\centering
\includegraphics[trim=0 0 0 30, clip,width=\columnwidth]{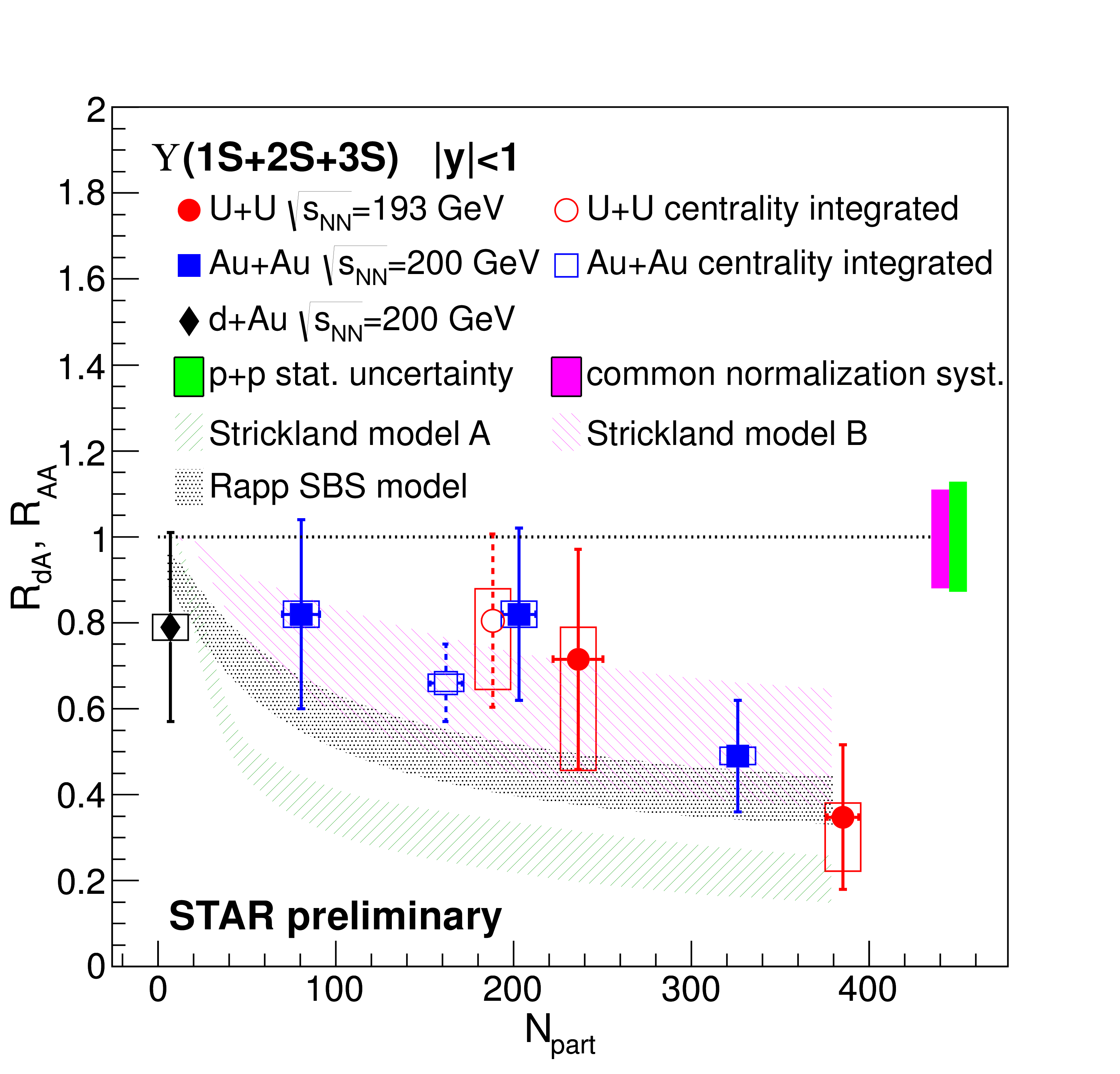}%
\caption{\Ups nuclear modification factor in \snn=200 GeV d+Au
  (rhombus) and Au+Au (squares)~\cite{Adamczyk:2013poh}, as well as
  new preliminary \snn=193 GeV U+U
  collisions (circles), compared to model calculations from Refs.~\cite{Strickland:2011aa,Emerick:2011xu}.}
\label{fig:Y_Raa}
\end{figure}

Nuclear modification factors of the $\Upsilon$(1S+2S+3S) in \snn{}=200 GeV d+Au, Au+Au and new \snn{}=193 GeV U+U
collisions are presented in Fig.~\ref{fig:Y_Raa} with respect to the number
of participant nucleons. The trend observed in Au+Au is generally continued in
the U+U data, with an $R_{\rm{AA}}$=0.35$\pm$0.17(stat.)$^{+0.03}_{-0.13}
$(syst.) measured for the 10\% most central U+U collisions. 
The model of Strickland and Bazow~\cite{Strickland:2011aa} incorporates lattice QCD
results on screening and broadening of bottomonium and the dynamical
propagation of the $\Upsilon$ meson in the colored medium. Assuming an
initial temperature between 428$<$T$<$443 \MeV, the scenario
with a potential based on heavy quark internal energy is consistent
with the observations, while the free energy based scenario is
disfavoured. The strong binding scenario in model proposed by Emerick,
Zhao, and Rapp~\cite{Emerick:2011xu}, which includes possible CNM effects in
addition, is also consistent with STAR results. It is to be noted, however,
that STAR observed a suppression beyond model predictions of \Ups in
$\sqrt{s_{NN}}=200$ GeV d+Au collisions~\cite{Adamczyk:2013poh}, which
indicates that further investigation of CNM effects is necessary. This
will be made possible by the upcoming year 2015 high-luminosity p+Au run.

\section{Future prospects}
\label{sec:outlook}

In recent years, two new detectors have been installed that facilitate
the ongoing heavy flavor program of STAR, the Muon Telescope Detector
(MTD) and the Heavy Flavor Tracker (HFT). 
Both the MTD and the HFT were operational and taking data with
dedicated trigger setups in the 2014 high-statistics Au+Au run.

\begin{figure}[t]
\centering
\includegraphics[width=\columnwidth]{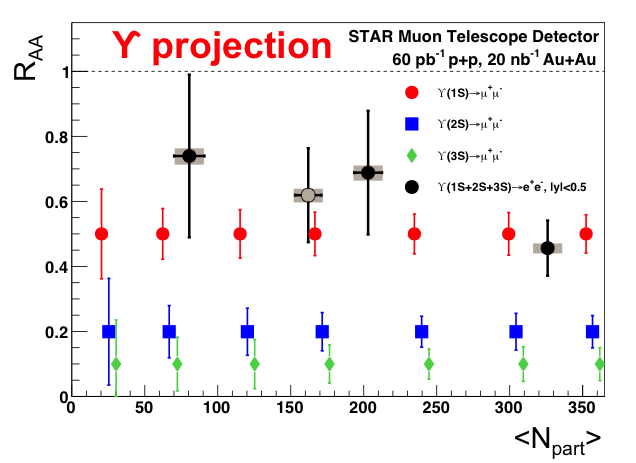}%
\caption{Anticipated statistical uncertainties on \Raa as a function of \Npart
measured in \snn{}=200 GeV Au+Au collisions the \Ups{}$\rightarrow${}$\mu^+\mu^-$ channel for
$\Ups(1S)$ (red), $\Ups(2S)$ (blue) and $\Ups(3S)$ (green)
separately. The $\Ups(1S+2S+3S)$ measured with the BEMC in the
electron channel (black) is plotted for comparison purposes.}
\label{fig:MTD_Yprojection}
\end{figure}

The MTD ~\cite{Ruan:2009ug} is a multi-gap resistive plate chamber that is located outside the STAR magnet, and covers 45\% of the $|\eta|${}$<$0.5 region.
It is capable of detecting muons with $\approx$90\%
efficiency up to \pT=20 GeV/$c$ along with good hadron
rejection power. MTD allows for precision quarkonium \Raa and
$v_2$ measurements.
Fig.~\ref{fig:MTD_Yprojection} shows the precision on \Raa versus \Npart
anticipated in the \Ups ($n$S){}$\rightarrow${}$\mu^+\mu^-$ channel for the
three states separately.

\begin{figure}[t]
\centering
\includegraphics[width=0.95\columnwidth]{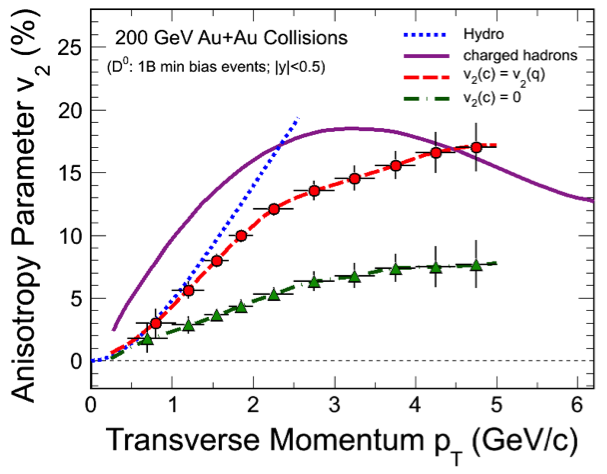}%
\caption{Anticipated \Dz $v_2$ statistical uncertainties versus \pT in
  \snn{}=200 GeV Au+Au
  collisions for the scenario when charm and light quarks flow
  similarly (red) and when charm has no flow (green).}
\label{fig:HFT_D0_v2projection}
\end{figure}

The HFT~\cite{Xu:2006dx} is a silicon microvertex detector system consisting of three
subdetectors. Two layers of silicon pixel detectors (PXL) are located
around the beampipe. The Intermediate Silicon Tracker (IST) is a
single layer of silicon pad detectors. The outermost layer is the
double-sided Silicon Strip Detector (SSD). The HFT makes it possible
to reconstruct a secondary vertex with 20 $\mu$m
precision. Note that the typical displacement of
\Dz$\rightarrow${}$\Km\pip$, $\Lambda_c^+${}$\rightarrow${}$p\Km\pip$ and B$^\pm${}$\rightarrow${}$\Jpsi$+X secondary
vertices are $c\tau${}$\approx$120, 60 and 500 $\mu$m,
respectively. Thus, through the topological reconstruction of
heavy flavor decays, HFT allows for precision measurements of open heavy
flavor production and flow, as well as for the separation of prompt and
non-prompt \Jpsi{} contributions. As an example,
Fig.~\ref{fig:HFT_D0_v2projection} shows the projected uncertainties of the \Dz $v_2$ measurements in \snn{}=200 GeV Au+Au
  collisions.

\section{Summary}
\label{sec:summary}

A selection of recent heavy flavor measurements has been presented
here. Total charm production cross sections from p+p \Dz
and \Ds measurements and Au+Au \Dz measurements at three different
centralities are consistent with \Nbin scaling, which supports the
picture that charm is mainly produced in initial hard processes. The
production of \Dz
in top RHIC energy central heavy ion collisions at \pT{}$>$2 GeV/$c$
shows a substantial suppression similar to light hadrons,
suggesting a strong charm--medium interaction, while the hump-like
structure at 1$<$\pT{}$<$2 GeV/$c$ suggests charm--light quark
coalescence. While \snn{}=200 GeV non-photonic electron measurements
exhibit both a strong suppression and substantial anisothropy $v_2$,
\snn{}=62.4 GeV
\Raa shows no trace of suppression, and \snn{}=62.4 and 39 GeV $v_2$
measurements are consistent with no flow, suggesting a radical change
in heavy flavor production and dynamics between \snn{}=62.4 and 200 GeV energies.

The \Jpsi \Raa also shows a strong suppression, but with a weak beam energy
dependence. The U+U minimum bias data are consistent with the
Au+Au measurements within errors. On the other
hand, the measured $v_2$ for \Jpsi is consistent with non-flow,
thus rendering the thermalized $c\overline{c}$ coalescence scenario unlikely.
The high-\pT \Jpsi production, less influenced by regeneration and CNM
effects, is also significantly suppressed in
\snn{}=200 GeV central Au+Au collisions, which is a clear signal of the
sQGP. This conclusion is further supported by the quantitatively
similar suppression of \Ups{}(1S), and that the \Ups{}(2S) and \Ups{}(3S)
states are consistent with a complete suppression~\cite{Adamczyk:2013poh}. Cold nuclear matter
effects, however, may play an important role in the case of \Ups. The
\Ups{}(1S+2S+3S) \Raa versus \Npart seems to follow a universal trend
in d+Au, Au+Au and U+U, which is consistent with an sQGP initial temperature 428$<$T$<$443 MeV, according to a model which assumes an internal energy based potential~\cite{Strickland:2011aa}.

\section*{Acknowledgements}
\label{sec:ack}
This work has been supported by the grant 13-02841S of the Czech
Science Foundation (GA\v{C}R), and by the MSMT grant
CZ.1.07/2.3.00/20.0207 of the European Social Fund (ESF) in the Czech
Republic: ``Education for Competitiveness Operational Programme''
(ECOP).

\end{document}